\documentclass[10pt,a4paper,notitlepage]{article}
\usepackage[utf8]{inputenc}
\usepackage[T1]{fontenc}
\usepackage{multicol} 
\usepackage[margin=0.75in]{geometry} 
\usepackage[english]{babel}
\usepackage{csquotes}
\usepackage{titlesec} 
\usepackage{float} 
\usepackage{makecell} 
\usepackage{tabularx}
\usepackage{amssymb} 
\usepackage{siunitx} 
\usepackage{amsmath} 
\titleformat*{\section}{\normalsize\bfseries} 
\titleformat*{\subsection}{\normalsize\bfseries} 
\titleformat*{\subsubsection}{\normalsize\bfseries} 
\usepackage{hyperref} 
\usepackage[backend=biber, sorting=none, style=nature]{biblatex} 
\usepackage{tcolorbox}
\usepackage{enumitem}

\begin{document}
\begin{center}
\Large 
\textbf{Geometric Deep Learning on Molecular Representations}\\

\medbreak \medbreak

\small {Kenneth Atz}$^{1, \dagger}$, {Francesca Grisoni}$^{2, 1, \dagger*}$, {Gisbert Schneider}$^{1, 3*}$ \medbreak
\footnotesize $^{1}$ETH Zurich, Dept. Chemistry and Applied Biosciences, RETHINK, Vladimir-Prelog-Weg 4, 8093 Zurich, Switzerland. \par 
\footnotesize $^{2}$Eindhoven University of Technology, Dept. Biomedical Engineering, Groene Loper 7, 5612AZ Eindhoven, Netherlands. \par 
\footnotesize $^{3}$ETH Singapore SEC Ltd, 1 CREATE Way, $\#$06-01 CREATE Tower, Singapore, Singapore. \par 
\footnotesize $\dagger$ these authors contributed equally to this work \par 
\footnotesize *f.grisoni@tue.nl, gisbert@ethz.ch \par
\end{center}

\begingroup 
\begin{abstract}
\noindent
Geometric deep learning (GDL), which is based on neural network architectures that incorporate and process symmetry information, has emerged as a recent paradigm in artificial intelligence. GDL bears particular promise in molecular modeling applications, in which various molecular representations with different symmetry properties and levels of abstraction exist. This review provides a structured and harmonized overview of molecular GDL, highlighting its applications in drug discovery, chemical synthesis prediction, and quantum chemistry. Emphasis is placed on the relevance of the learned molecular features and their complementarity to well-established molecular descriptors. This review provides an overview of current challenges and opportunities, and presents a forecast of the future of GDL for molecular sciences. 

\end{abstract}
\endgroup

\begin{multicols}{2}
\raggedcolumns

\section{Introduction}

\begin{figure}[H]
\centering
\label{figmolrep}\includegraphics[width=8cm]{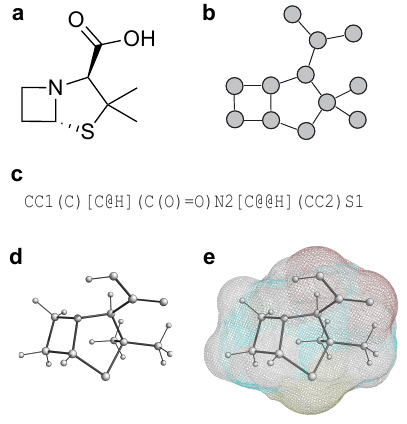}
\caption{\textit{Exemplary molecular representations for a selected molecule (i.e., the penam substructure of penicillin)}\\ 
\textbf{a.} Two-dimensional (2D) depiction (Kekulé structure).\\ \textbf{b.} Molecular graph (2D), composed of vertices (atoms) and edges (bonds). \\ \textbf{c.} SMILES string \cite{weininger1988smiles}, in which atom type, bond type and connectivity are specified by alphanumerical characters.\\ \textbf{d.} Three-dimensional (3D) graph, composed of vertices (atoms), their position ($x$, $y$, $z$ coordinates) in 3D space, and edges (bonds). \\ \textbf{e.} Molecular surface represented as a mesh colored according to the respective atom types.}
\end{figure}

Recent advances in deep learning, which is an instance of artificial intelligence (AI) based on neural networks \cite{lecun2015deep,schmidhuber2015deep}, have led to numerous applications in the molecular sciences, \textit{e.g.}, in drug discovery \cite{gawehn2016deep,jimenez2021artificial}, quantum chemistry \cite{gilmer2017neural}, and structural biology \cite{jumper2021highly,baek2021accurate}. Two characteristics of deep learning render it particularly promising when applied to molecules. First, deep learning methods can cope with "unstructured" data representations, such as text sequences \cite{vaswani2017attention, brown2020language}, speech signals \cite{hinton2012deep, mikolov2011strategies}, images \cite{krizhevsky2017imagenet,farabet2012learning,tompson2014joint}, and graphs \cite{bronstein2017geometric,monti2019fake}. This ability is particularly useful for molecular systems, for which chemists have developed many models (\textit{i.e.}, "molecular representations") that capture molecular properties at varying levels of abstraction (Figure \ref{figmolrep}). The second key characteristic is that deep learning can perform feature extraction (or feature learning) from the input data, that is, produce data-driven features from the input data without the need for manual intervention. These two characteristics are promising for deep learning as a complement to “classical” machine learning applications (\textit{e.g.}, Quantitative Structure-Activity Relationship [QSAR]), in which molecular features (\textit{i.e.}, "molecular descriptors" \cite{todeschini2009molecular}) are encoded \textit{a priori} with rule-based algorithms. The capability to learn from unstructured data and obtain data-driven molecular features has led to unprecedented applications of AI in the molecular sciences. 

One of the most promising advances in deep learning is geometric deep learning (GDL). \textit{Geometric deep learning} is an umbrella term encompassing emerging techniques which generalize neural networks to Euclidean and non-Euclidean domains, such as graphs, manifolds, meshes, or string representations \cite{bronstein2017geometric}. In general, GDL encompasses approaches that incorporate a geometric prior, \textit{i.e.}, information on the structure space and symmetry properties of the input variables. Such a geometric prior is leveraged to improve the quality of the information captured by the model. Although GDL has been increasingly applied to molecular modeling \cite{gainza2020deciphering,segler2018generating,gilmer2017neural}, its full potential in the field is still untapped.

The aim of this review is to (i) provide a structured and harmonized overview of the applications of GDL on molecular systems, (ii) delineate the main research directions in the field, and (iii) provide a forecast of the future impact of GDL. Three fields of application are highlighted, namely drug discovery, quantum chemistry, and computer-aided synthesis planning (CASP), with particular attention to the data-driven molecular features learned by GDL methods. A glossary of selected terms can be found in Box 1.

\section{Principles of geometric deep learning}

The term \textit{geometric} \textit{deep} \textit{learning} was coined in 2017 \cite{bronstein2017geometric}. Although GDL was originally used for methods applied to non-Euclidean data \cite{bronstein2017geometric}, it now extends to all deep learning methods that incorporate geometric priors \cite{bronstein2021geometric}, that is, information about the structure and symmetry of the system of interest. Symmetry is a crucial concept in GDL, as it encompasses the properties of the system with respect to manipulations (transformations), such as translation, reflection, rotation, scaling, or permutation (Box 2). 

Symmetry is often recast in terms of \textit{invariance} and \textit{equivariance} to express the behavior of any mathematical function with respect to a transformation $\mathcal{T}$ (\textit{e.g.} rotation, translation, reflection or permutation) of an acting symmetry group \cite{marsden1974reduction}. Here, the mathematical function is a neural network $\mathcal{F}$ applied to a given molecular input $\mathcal{X}$. $\mathcal{F}(\mathcal{X})$ can therein transform equivariantly, invariantly or neither with respect to $\mathcal{T}$, as described below:  

\begin{itemize}
    \item \textit{Equivariance.} A neural network $\mathcal{F}$ applied to an input $\mathcal{X}$ is \textit{equivariant} to a transformation $\mathcal{T}$ if the transformation of the input $\mathcal{X}$ commutes with the transformation of $\mathcal{F}(\mathcal{X})$, \textit{via} a transformation $\mathcal{T}'$ of the same symmetry group, such that: $\mathcal{F}(\mathcal{T}(\mathcal{X})) = \mathcal{T}' \mathcal{F}(\mathcal{X})$. Neural networks are therefore equivariant to the actions of a symmetry group on their inputs if and only if each layer of the network “equivalently" transforms under any transformation of that group. 
    
    \item \textit{Invariance.} Invariance is a special case of equivariance, where $\mathcal{F}(\mathcal{X})$ is invariant to $\mathcal{T}$ if $\mathcal{T}'$ is the trivial group action (\textit{i.e}., identity): $\mathcal{F}(\mathcal{T}(\mathcal{X})) = \mathcal{T}' \mathcal{F}(\mathcal{X}) = \mathcal{F}(\mathcal{X})$. 
    
    \item $\mathcal{F}(\mathcal{X})$ is neither equivariant nor invariant to $\mathcal{T}$ when the transformation of the input $\mathcal{X}$ does not commute with the transformation of $\mathcal{F}(\mathcal{X})$: $\mathcal{F}(\mathcal{T}(\mathcal{X})) \neq \mathcal{T}' \mathcal{F}(\mathcal{X})$.
\end{itemize}

The symmetry properties of a neural network architecture vary depending on the network type and the symmetry group of interest and are individually discussed in the following sections. Readers can find an in-depth treatment of equivariance and group equivariant layers in neural networks elsewhere \cite{cohen2016group,cohen2016steerable,cohen2018spherical,kondor2018generalization}.

The concept of equivariance and invariance can also be used in reference to the molecular features obtained from a given molecular representation ($\mathcal{X}$), depending on their behaviour when a transformation is applied to $\mathcal{X}$. For instance, many molecular descriptors are invariant to the rotation and translation of the molecular representation by design \cite{todeschini2009molecular}, \textit{e.g.,} the Moriguchi octanol-water partitioning coefficient \cite{moriguchi1992simple}, which relies only on the occurrence of specific molecular substructures for calculation. The symmetry properties of molecular features extracted by a neural network depend on both the symmetry properties of the input molecular representation and of the utilized neural network. 

Many relevant molecular properties (\textit{e.g.}, equilibrium energies, atomic charges, or physicochemical properties such as permeability, lipophilicity or solubility) are invariant to certain symmetry operations (Box 2). In many tasks in chemistry, it is thus desirable to design neural networks that transform equivariantly under the actions of pre-defined symmetry groups. Exceptions occur if the targeted property changes upon a symmetry transformation of the molecules (\textit{e.g}., chiral properties which change under inversion of the molecule, or vector properties which change under rotation of the molecule). In such cases, the inductive bias (learning bias) of equivariant neural networks would not allow for the differentiation of symmetry-transformed molecules. 

While neural networks can be considered as universal function approximators \cite{cybenko1989approximation}, incorporating prior knowledge such as reasonable geometric information (geometric priors) has evolved as a core design principle of neural network modeling \cite{bronstein2021geometric}. By incorporating geometric priors, GDL allows to increase the quality of the model and bypasses several bottlenecks related to the need to force the data into Euclidean geometries (\textit{e.g.}, by feature engineering). Moreover, GDL provides novel modeling opportunities, such as data augmentation in low data regimes \cite{tetko2020state,skinnider2021chemical}. 

\begin{tcolorbox}[float*=t, floatplacement=t, width=\textwidth]
\subsection*{Box 1: Glossary of selected terms}

\textbf{CoMFA and CoMSIA.} Comparative Molecular Field Analysis (CoMFA) \cite{cramer1988comparative} and Comparative Molecular Similarity Indices Analysis (CoMSIA) \cite{klebe1998comparative} are popular 3D QSAR methods developed in the 1980s and 1990s, in which three-dimensional grids are used to capture the distributions of molecular features (\textit{e.g}., steric, hydrophobic, and electrostatic properties). The obtained molecular descriptors serve as inputs to a regression model for quantitative bioactivity prediction.

\textbf{Convolution}. Operation within a neural network that transforms a feature space into a new feature space and thereby captures the local information found in the data. Convolutions were first introduced for pixels in images \cite{lecun1995convolutional,lecun1998gradient} but the term "convolution" is now used for neural network architectures covering a variety of data structures such as graphs, point clouds, spheres, grids, or manifolds.

\textbf{Density Functional Theory (DFT)}. A quantum mechanical modeling approach used to investigate the electronic structure of molecules. 

\textbf{Data augmentation}. Artificial increase of the data volume available for model training, often achieved by leveraging symmetrical properties of the input data which are not captured by the model (\textit{e.g.}, rotation or permutation).

\textbf{Feature}. An individually measurable or computationally obtainable characteristic of a given sample (\textit{e.g.}, molecule), in the form of a scalar. In this review, the term refers to a numeric value characterizing a molecule. Such molecular features can be computed with rule-based algorithms ("molecular descriptors") or generated automatically by deep learning from a molecular representation ("hidden" or "learned" features). 

\textbf{Geometric prior.} An inductive bias incorporating information on the symmetric nature of the system of interest into the neural network architecture. Also known as \textit{symmetry prior}.

\textbf{Inductive bias}. Set of assumptions that a learning algorithm (\textit{e.g.}, a neural network) uses to learn the target function and to make predictions on previously unseen data points.

\textbf{One-hot encoding}. Method for representing categorical variables as numerical arrays by obtaining a binary variable (0, 1) for each category. It is often used to convert sequences (\textit{e.g.}, SMILES strings) into numerical matrices, suitable as inputs and/or outputs of deep learning models (\textit{e.g.}, chemical language models).

\textbf{Quantitative Structure-Activity Relationship (QSAR).} Machine learning techniques aimed at finding an empirical relationship between the molecular structure (usually encoded as molecular descriptors) and experimentally determined molecular properties, such as pharmacological activity or toxicity.

\textbf{Reinforcement learning}. A technique used to steer the output of a machine learning algorithm toward user-defined regions of optimality \textit{via} a predefined reward function \cite{sutton2018reinforcement}.

\textbf{Transfer learning}. Transfer of knowledge from an existing deep learning model to a related task for which fewer training samples are available \cite{pan2009survey}.

\textbf{Unstructured data}. Data that are not arranged as vectors of (typically handcrafted) features. Examples of unstructured data include graphs, images, and meshes. Molecular representations are typically unstructured, whereas numerical molecular descriptors (\textit{e.g}., molecular properties, molecular "fingerprints") are examples of structured data. 

\textbf{Voxel}. Element of a regularly spaced, 3D grid (equivalent to a pixel in 2D space). 

\end{tcolorbox}

\newcolumntype{s}{>{\hsize=.5\hsize}X}
\newcolumntype{m}{>{\hsize=.7\hsize}X}

\begin{table*}[!htbp]
\caption{\textit{Summary of selected geometric deep learning (GDL) approaches for molecular modeling}. For each approach, the utilized molecular representation(s) and selected applications are reported. 1D, one-dimensional; 2D, two-dimensional; 3D, three-dimensional.}
\centering
\label{tab:summary}
\begin{tabularx}{\textwidth}{smX}
\Xhline{3\arrayrulewidth}
\textbf{GDL approach} & \textbf{Molecular representation(s)} & \textbf{Applications} \\ [0.5ex] 
\Xhline{3\arrayrulewidth}
Graph neural networks (GNNs)& 2D and 3D molecular graph, and 3D point cloud. & Molecular property prediction in drug discovery  \cite{feinberg2018potentialnet,jimenez2021coloring} and in quantum chemistry for energies \cite{miller2020relevance,anderson2019cormorant,satorras2021n}, forces \cite{fuchs2020se,satorras2021n,schutt2021equivariant} and wave-functions \cite{unke2021se3equivariant}, CASP \cite{coley2019graph,jin2017predicting}, and generative molecular design \cite{zhou2019optimization, jin2018junction}. \\ 
\Xhline{2\arrayrulewidth}
3D convolutional neural networks (3D CNNs)& 3D grid. & Structure-based drug design and property prediction \cite{jimenez2018k, ragoza2017protein}. \\
\Xhline{2\arrayrulewidth}
Mesh convolutional neural networks (geodesic CNNs or 3D GNNs)& Surface (mesh) encoded as a 2D grid or 3D graph. & Protein-protein interaction prediction and ligand-pocket fingerprinting \cite{gainza2020deciphering}. \\
\Xhline{2\arrayrulewidth}
Recurrent neural networks (RNNs) & String notation (1D grid). & Generative molecular design \cite{segler2018generating,grisoni2018designing}, synthesis planning \cite{schwaller2018found}, protein structure prediction \cite{senior2019protein} and prediction of properties in drug discovery \cite{wang2020optimizing,zheng2020predicting}. \\
\Xhline{2\arrayrulewidth}
Transformers & String notation encoded as a graph. & Synthesis planning \cite{schwaller2019molecular}, prediction of reaction yields \cite{schwaller2021prediction}, generative molecular design \cite{grechishnikova2021transformer}, prediction of properties in drug discovery \cite{morris_2020_transformer}, and protein structure prediction \cite{jumper2021highly,baek2021accurate}. \\
\Xhline{2\arrayrulewidth}
\end{tabularx}
\end{table*}

\section{Molecular GDL}

The application of GDL to molecular systems is challenging, in part because there are multiple valid ways of representing the same molecular entity. Molecular representations\footnote{Note that in this review the term "representation" is used solely to denote human-made models of molecules (\textit{e.g.}, molecular graphs, 3D conformers, SMILES strings). To avoid confusion with other usages of the word "representation" in deep learning, we will use the term "feature" whenever referring to any numerical description of molecules, obtained either with rule-based algorithms (molecular descriptors) or learned (extracted) by neural networks.} can be categorized based on their different levels of abstraction and the physicochemical and geometrical aspects they capture. Importantly, all of these representations are models of the same reality and are thus \textit{"suitable for some purposes, not for others"} \cite{hoffmann1991representation}. GDL provides the opportunity to experiment with different representations of the same molecule and leverages their intrinsic geometrical features to increase the quality of the model. Moreover, GDL has repeatedly proven useful in providing insights into relevant molecular properties for the task at hand, thanks to its feature extraction (feature learning) capabilities. In the following sections, we delineate the most prevalent molecular GDL approaches and their applications in chemistry, grouped according to the respective molecular representations used for deep learning: molecular graphs, grids, strings, and surfaces.

\begin{tcolorbox}[float*=t, floatplacement=t, width=\textwidth]
\subsection*{Box 2: Euclidean symmetries in molecular systems}
Molecular systems (and three-dimensional representations thereof) can be considered as objects in Euclidean space. In such a space, one can apply several symmetry operations (transformations) that are (i) performed with respect to three symmetry elements (\textit{i.e.}, line, plane, point), and (ii) rigid, that is, they preserve the Euclidean distance between all pairs of atoms (\textit{i.e.}, isometry). The Euclidean transformations are as follows:
\begin{itemize}[topsep=0.5pt, partopsep=0.5pt, itemsep=-0.3pt]
  \setlength\itemsep{0.003em}
  \item \textit{Rotation}. Movement of an object with respect to the radial orientation to a given point.
  \item \textit{Translation}. Movement of every point of an object by the same distance in a given direction.
  \item \textit{Reflection}. Mapping of an object to itself through a point (inversion), a line or a plane (mirroring).
\end{itemize}

All three transformations and their arbitrary finite combinations are included in  the \textit{Euclidean group} [E(3)]. The \textit{special Euclidean group} [SE(3)] comprises only translations and rotations.

Molecules are always symmetric in the SE(3) group, \textit{i.e.}, their intrinsic properties (e.g., biological and physicochemical properties, and equilibrium energy) are invariant to coordinate rotation and translation, and combinations thereof. Several molecules are chiral, that is, some of their (chiral) properties depend on the absolute configuration of their stereogenic centers, and are thus non-invariant to molecule reflection. Chirality plays a key role in chemical biology; relevant examples of chiral molecules are DNA, and several drugs whose enantiomers exhibit markedly different pharmacological and toxicological properties \cite{nguyen2006chiral}.

\begin{center}
\includegraphics[width=12cm]{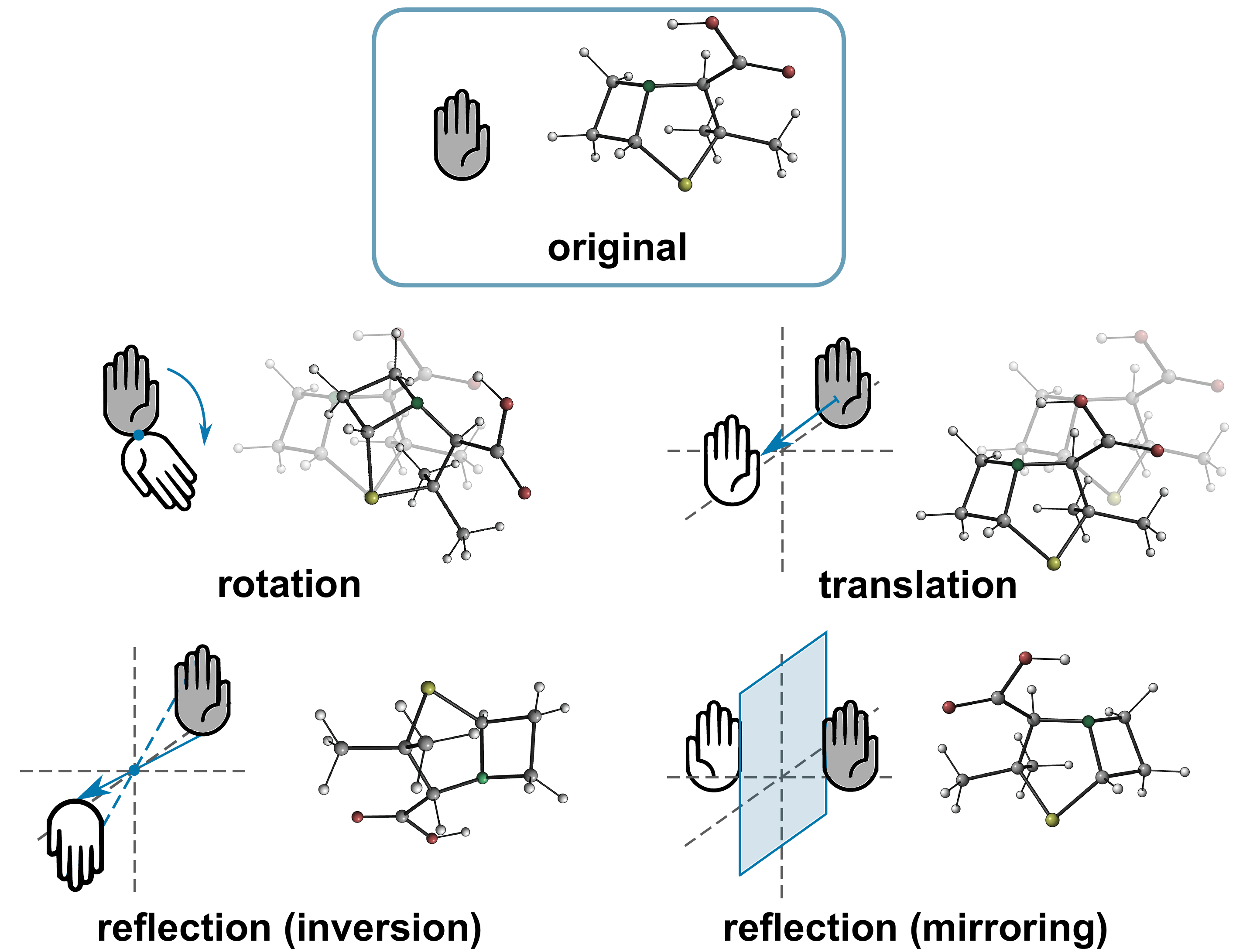}
\end{center}
\end{tcolorbox}

\subsection{Learning on molecular graphs} \label{graphmpnn}

\begin{figure*}[th!]
\centering
\includegraphics[width=17cm]{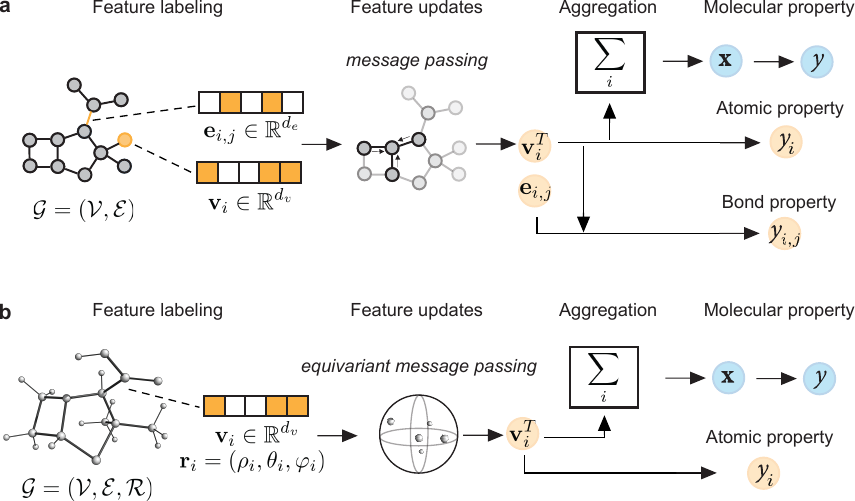} 
\caption{\textit{Deep learning on molecular graphs.}\\
\textbf{a.} Message passing graph neural networks applied to two-dimensional (2D) molecular graphs: 2D molecular graph $\mathcal{G} = (\mathcal{V}, \mathcal{E})$ with its labeled vertex (atom) features ($\textbf{v}_{i} \in \mathbb{R}^{d_v}$), and edge (bond) features ($\textbf{e}_{ij} \in \mathbb{R}^{d_e}$). Vertex features are updated by iterative message passing for a defined number of time steps $T$ across each pair of vertices ${v}_{i}$ and ${v}_{j}$, connected \textit{via} an edge $e_{j,i}$. After the last message passing convolution, the final vertex $\textbf{v}_{i}^{t}$ can be (i) mapped to a bond ($y_{ij}$) or atom ($y_{i}$) property, or (ii) aggregated to form molecular features (that can be mapped to a molecular property $y$).\\ 
\textbf{b.} E(3)-equivariant message passing graph neural networks applied to three-dimensional (3D) molecular graphs: 3D graphs $\mathcal{G}_3 = (\mathcal{V}, \mathcal{E}, \mathcal{R})$ that are labeled with atom features ($\textbf{v}_{i} \in \mathbb{R}^{d_v}$), their absolute coordinates in 3D space ($\textbf{r}_{i} \in \mathbb{R}^{3}$) and their edge features ($\textbf{e}_{ij} \in \mathbb{R}^{d_e}$). Iterative spherical convolutions are used to obtain data-driven atomic features ($\textbf{v}_{i}^{t}$), which can be mapped to atomic properties or aggregated, and mapped to molecular properties (${y}_{i}$ and ${y}$, respectively).
}
\label{message}
\end{figure*}

\subsubsection{Molecular graphs}

Graphs are among the most intuitive ways to represent molecular structures \cite{kipf2016semi}. Any molecule can be thought of as a mathematical graph $\mathcal{G} = (\mathcal{V}, \mathcal{E})$, whose vertices ($\textbf{v}_{i} \in \mathcal{V}$) represent atoms, and whose edges ($\textbf{e}_{i,j} \in \mathcal{E}$) constitute their connection (Figure \ref{graphmpnn}). In many deep learning applications, molecular graphs can be further characterized by a set of vertex and edge features. 

\subsubsection{Graph neural networks}

Deep learning methods devoted to handling graphs as input are commonly referred to as graph neural networks (GNNs). When applied to molecules, GNNs allow for feature extraction by progressively aggregating information from atoms and their molecular environments (Figure ~\ref{message}a, \cite{battaglia2016interaction, battaglia2018relational}). Different architectures of GNNs have been introduced \cite{zhou2020graph}, the most popular of which fall under the umbrella term of \textit{message passing neural networks} \cite{geerts2020let, duvenaud2015convolutional, gilmer2017neural}. Such networks iteratively update the vertex features of the \textit{l}-th network layer ($\textbf{v}_{i}^{l} \rightarrow \textbf{v}_{i}^{l+1}$) \textit{via} graph convolutional operations, employing at least two learnable functions $\psi$ and $\phi$, and a local permutation-invariant aggregation operator (\textit{\textit{e.g.},} sum): $\textbf{v}_{i}^{l+1} = \phi\left(\textbf{v}_{i}^{l}, \bigoplus_{j \in \mathcal{N}(i)}\psi\left(\textbf{v}_{i}^{l}, \textbf{v}_{j}^{l}\right)\right)$. 

Since their introduction as a means to predict quantum chemical properties of small molecules at the density functional theory (DFT) level \cite{gilmer2017neural}, GNNs have found many applications in quantum chemistry \cite{klicpera2020directional,mxmnet2020molecular,withnall2020building,tang2020self,liu2021spherical}, drug discovery \cite{stokes2020deep,feinberg2018potentialnet,torng2019graph}, CASP \cite{somnath2020learning}, and molecular property prediction \cite{li2017learning,liu2019chemi}. When applied to quantum chemistry tasks, GNNs often use E(3)-invariant 3D information by including radial and angular information into the edge features of the graph \cite{unke2019physnet, klicpera2020directional,mxmnet2020molecular,schutt2021equivariant,liu2021spherical}, thereby improving the prediction accuracy of quantum chemical forces and energies for equilibrium and non-equilibrium molecular conformations, as in the case of SchNet \cite{schutt2018schnet,schutt2017quantum} and PaiNN \cite{schutt2021equivariant}. SchNet-like architectures were used to predict quantum mechanical wave-functions in the form of Hartree-Fock and DFT density matrices \cite{schutt2019unifying}, and differences in quantum properties obtained by DFT and coupled cluster level-of-theory calculations \cite{bogojeski2020quantum}. 

GNNs for molecular property prediction have been shown to outperform human-engineered molecular descriptors for several biologically relevant properties \cite{yang2019analyzing}. Although including 3D information into molecular graphs generally improved the prediction of drug-relevant properties, no marked difference was observed between using a single or multiple molecular conformers for network training \cite{axelrod2020molecular}. Because of their natural connection with molecular representations, GNNs seem particularly suitable in the context of explainable AI (XAI) \cite{jimenez2020drug}, where they have been used to interpret models predicting molecular properties of preclinical relevance \cite{jimenez2021coloring} and quantum chemical properties \cite{schnake2020xai}. 

GNNs have been used for \textit{de novo} molecule generation \cite{battaglia2018learning, simonovsky2018graphvae,de2018molgan,zhou2019optimization}, for example by performing vertex and edge addition from an initial vertex \cite{battaglia2018learning} (Figure ~\ref{message}b). GNNs have also been combined with variational autoencoders \cite{jin2018junction,de2018molgan,simonovsky2018graphvae,flam2021mpgvae} and reinforcement learning \cite{you2018graph,jin2020multi,zhou2019optimization}. 
Finally, GNNs have been applied to CASP \cite{somnath2020learning,coley2019graph,lei2017deriving}; however, the current approaches are limited to reactions in which one bond is removed between the products and the reactants.

\subsubsection{Equivariant message passing}

A recent area of development of graph-based methods are SE(3)- and E(3)-equivariant GNNs (\textit{equivariant message passing networks}) which deal with the absolute coordinate systems of 3D graphs \cite{thomas2018tensor,smidt2021finding} (Figure \ref{message}b). Thus, these networks may be particularly well-suited to be applied to 3D molecular representations. Such networks exploit Euclidean symmetries of the system (Box 2).

3D molecular graphs $\mathcal{G}_{3D} = (\mathcal{V}, \mathcal{E}, \mathcal{R})$, in addition to their vertex and edge features ($\textbf{v}_{i} \in \mathcal{V}$ and $\textbf{e}_{ij} \in \mathcal{E}$, respectively), also encode information on the vertex position in a 3D coordinate system ($\textbf{r}_{i} \in \mathcal{R}$). By employing E(3)- \cite{satorras2021n} and SE(3)-equivariant \cite{thomas2018tensor} convolutions, such networks have shown high accuracy for predicting several quantum chemical properties such as energies \cite{anderson2019cormorant,smidt2020euclidean,miller2020relevance,fuchs2020se,hutchinson2020lietransformer,schutt2021equivariant,unke2021spookynet}, interatomic potentials for molecular dynamics simulations \cite{fuchs2020se,batzner2021se,satorras2021n}, and wave-functions \cite{unke2021se3equivariant}. SE(3) equivariant neural networks do not commute with reflections of the input  (\textit{i.e.} non-equivariant to reflections), and thereby enable SE(3) equivariant models to distinguish between stereoisomers of chiral molecules including enantiomers \cite{thomas2018tensor}. E(3) equivariant neural networks on the other side transform equivariantly with refelctions, which allows E(3) equivariant models only to distinguish between diastereomers and not eneantiomers. SE(3) neural networks are computationally expensive due to their use of \textit{spherical harmonics} \cite{muller2006spherical} and \textit{Wigner D-functions} \cite{dray1986unified} to compute learnable weight kernels. E(3)-equivariant neural networks are computationally more efficient and have shown to perform equal to, or better than, SE(3)-equivariant networks, \textit{e.g.}, for the modeling of quantum chemical properties and dynamic systems \cite{satorras2021n}.
Equivariant message passing networks have been applied to predict the quantum mechanical wave-function of nuclei and electron-based representations in an end-to-end fashion \cite{hermann2020deep,pfau2020ab,choo2020fermionic}. However, such networks are currently limited to small molecular systems because of the large size of the learned matrices, which scale quadratically with the number of electrons in the system.

\subsection{Learning on grids}

Grids capture the properties of a system at regularly spaced intervals. Based on the number of dimensions included in the system, grids can be 1D (\textit{e.g.}, sequences), 2D (\textit{e.g.}, RGB images), 3D (\textit{e.g.}, cubic lattices), or higher-dimensional. Grids are defined by a Euclidean geometry and can be considered as a graph with a special adjacency, where (i) the vertices have a fixed ordering that is defined by the spatial dimensions of the grid, and (ii) each vertex has an identical number of adjacent edges and is therefore indistinguishable from all other vertices structure-wise \cite{bronstein2021geometric}. These two properties render local convolutions applied to a grid inherently permutation invariant, and provide a strong geometric prior for translation invariance (\textit{e.g.} by weight sharing in convolutions). These grid properties have critically determined the success of convolutional neural networks (CNNs), \textit{e.g.}, in computer vision \cite{lecun1998gradient,lecun1995convolutional}, natural language processing \cite{hochreiter1997long,brown2020language}, and speech recognition \cite{hinton2012deep,mikolov2011strategies}.

\subsubsection{Molecular grids}
Molecules can be represented as grids in different ways. 2D grids (\textit{e.g.}, molecular structure drawings) are generally more useful for visualization rather than prediction, with few exceptions \cite{rajan2020decimer}. Analogous with some popular pre-deep learning approaches, for example Comparative molecular field analysis (CoMFA) \cite{cramer1988comparative}, and comparative molecular similarity indices analysis (CoMSIA) \cite{klebe1998comparative}, 3D grids are often used to capture the spatial distribution of the properties within one (or more) molecular conformer. Such representations are then used as inputs to the 3D CNNs. 3D CNNs are characterized by a greater resource efficiency than equivariant GNNs, which until now have mainly been applied to molecules with fewer than approximately 1000 atoms. Thus, 3D CNNs have often been the method of choice when the protein structure has to be considered, \textit{e.g.}, for protein-ligand binding affinity prediction \cite{jimenez2018k,ragoza2017protein,li2019deepatom,karimi2019deepaffinity,jimenez2019deltadelta}, or active site recognition \cite{jimenez2017deepsite}.

\subsection{Learning on molecular surfaces}

Molecular surfaces can be defined by the surface enclosing the 3D structure of a molecule at a certain distance from each atom center. Each point on such a continuous surface can be further characterized by its chemical (\textit{e.g.}, hydrophobic, electrostatic) and geometric features (\textit{e.g.}, local shape, curvature). From a geometrical perspective, molecular surfaces are considered as 3D meshes, \textit{i.e.}, a set of polygons (faces) that describe how the mesh coordinates exist in the 3D space \cite{ahmed2018survey}. Their vertices can be represented by a 2D grid structure (where four vertices on the mesh define a pixel) or by a 3D graph structure. The grid- and graph-based structures of meshes enable applications of 2D CNNs, geodesic CNNs and GNNs to learn on mesh-based molecular surfaces. Recently, geodesic (2D) CNNs have been applied to learn on mesh-based representations of protein surfaces to predict protein-protein interactions and recognize corresponding binding sites \cite{gainza2020deciphering}. This approach generated data-driven fingerprints that are relevant for specific biomolecular interactions. Approaches like 2D CNNs applied to meshes come with certain limitations, such as the need for rotational data augmentation (due to their non-equivariance to rotations) and for enforcing a homogeneous mesh resolution  (\textit{i.e.}, uniform spacing of all the points in the mesh). Recently introduced GNNs for mesh-based representations have been shown to incorporate rotational equivariance into their network architecture and allow for heterogeneous mesh resolution \cite{pfaff2020learning}. Such GNNs are computationally efficient and have potential for modeling macromolecular structures; however, they have not yet found applications to molecular systems. Other studies have used 3D voxel-based surface representations of (macro)molecules as inputs to 3D CNNs, \textit{\textit{e.g.},} for protein-ligand affinity \cite{liu2021octsurf} and protein binding-site \cite{mylonas2020deepsurf} prediction.

\subsection{Learning on string representations}

\subsubsection{Molecular strings}

Molecules can be represented as molecular strings, \textit{i.e.}, linear sequences of alphanumeric symbols. Molecular strings were originally developed as manual ciphering tools to complement systematic chemical nomenclature \cite{barnard2003representation,wiswesser1985historic} and later became suitable for data storage and retrieval. Some of the most popular string-based representations are the Wiswesser Line Notation \cite{wln_1952}, the Sybyl line notation \cite{ash1997sybyl}, the International Chemical Identifier (InChI) \cite{heller2013inchi}, Hierarchical Editing Language for Macromolecules \cite{zhang_2012_helm}, and the Simplified Molecular Input Line Entry System (SMILES) \cite{weininger1988smiles}. 

Each type of linear representation can be considered as a "chemical language." In fact, such notations possess a defined syntax, \textit{i.e.}, not all possible combinations of alphanumerical characters will lead to a “chemically valid” molecule. Furthermore, these notations possess semantic properties: depending on how the elements of the string are combined, the corresponding molecule will have different physicochemical and biological properties. These characteristics make it possible to extend the deep learning methods developed for language and sequence modeling to the analysis of molecular strings for "chemical language modeling" \cite{ozturk2020exploring,cadeddu_2014_language}.

\begin{figure*}[th!]
\centering
\includegraphics[width=17cm]{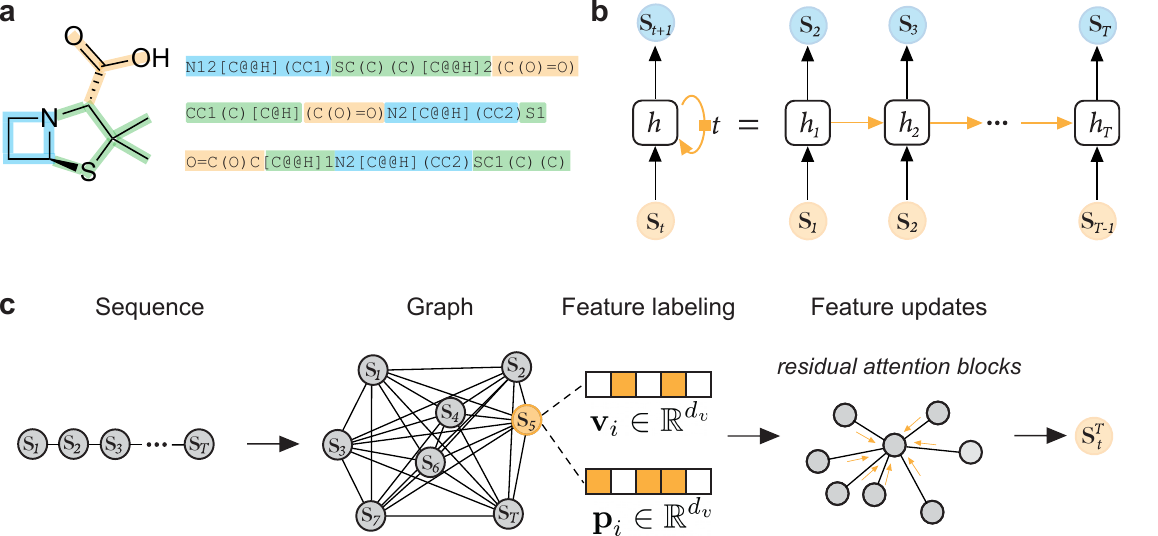}
\caption{\textit{Chemical language modeling}. \\ \textbf{a.} SMILES strings, in which atom types are represented by their element symbols, and bond types and branching are indicated by other predefined alphanumeric symbols. For each molecule, \textit{via} the SMILES algorithm a string of $T$ symbols ("tokens") is obtained ($\textbf{s} = \{{s}_1, {s}_2, \ldots, {s}_T\}$), which encodes the molecular connectivity, herein illustrated \textit{via} the color that indicates the corresponding atomic position in the graph (\textit{left}) and string (\textit{right}). A molecule can be encoded \textit{via} different SMILES strings depending on the chosen starting atom. Three random permutations incorporating identical molecular information are presented.\\ \textbf{b.} Recurrent neural networks, at any sequence position \textit{t}, learn to predict the next token ${s}_{t+1}$ of a sequence $\textbf{s}$ given the current sequence ($\{{s}_1, {s}_2, \ldots, {s}_t\}$) and hidden state ${h}_{t}$.\\ \textbf{c.} Transformer-based language models, in which the input sequence is structured as a graph. Vertices are featurized according to their token identity (\textit{e.g., } \textit{via} token embedding, $\textbf{v}_{i} \in \mathbb{R}^{d_v}$) and their position in the sequence (\textit{e.g.,} \textit{via} sinusoidal positional encoding, $\textbf{p}_{i} \in \mathbb{R}^{d_v}$). During transformer learning, the vertices are updated \textit{via} residual attention blocks. After passing $T$ attention layers, an individual feature representation $\textbf{s}_{t}^{T}$ for each token is obtained.}
\label{language}
\end{figure*}

SMILES strings -- in which letters are used to represent atoms, and symbols and numbers are used to encode bond types, connectivity, branching, and stereochemistry (Figure ~\ref{language}a) -- have become the most frequently employed data representation method for sequence-based deep learning \cite{segler2018generating,schwaller2018found}. Whereas several other string representations have been tested in combination with deep learning, \textit{\textit{e.g.},} InChI \cite{gomez2018automatic}, DeepSMILES \cite{deepsmiles}, and self-referencing embedded strings (SELFIES) \cite{krenn2020self}, SMILES remains the \textit{de facto} representation of choice for chemical language modeling \cite{skinnider2021chemical}. The following text introduces the most prominent chemical language modeling methods, along with selected examples of their application to chemistry.

\subsubsection{Chemical language models}

Chemical language models are machine learning methods that can handle molecular sequences as inputs and/or outputs. The most common algorithms for chemical language modeling are \textit{Recurrent neural networks} (RNNs) and \textit{Transformers}:
\begin{itemize}
  \item \textit{RNNs} (Figure ~\ref{language}b) \cite{rumelhart1985learning} are neural networks that process sequence data as \textit{Euclidean} structures, usually \textit{via} one-hot-encoding. RNNs model a dynamic system in which the hidden state (${h}_t$) of the network at any \textit{t}-th time point (\textit{i.e.}, at any \textit{t}-th position in the sequence) depends on both the current observation (${s}_t$) and the previous hidden state (${h}_{t-1}$). RNNs can process sequence inputs of arbitrary lengths and provide outputs of arbitrary lengths. RNNs are often used in an "auto-regressive" fashion, \textit{i.e.}, to predict the probability distribution over the next possible elements (tokens) at the time step $t+1$, given the current hidden state (${h}_t$) and the preceding portions of the sequence. Several RNN architectures have been proposed to solve the gradient vanishing or exploding problems of "vanilla" RNNs \cite{hochreiter1998vanishing, pascanu2013difficulty}, such as long short-term memory \cite{hochreiter1997long} and gated recurrent units \cite{chung2014empirical}.
  
  \item \textit{Transformers} (Figure ~\ref{language}c) process sequence data as \textit{non-Euclidean} structures, by encoding sequences as either (i) a fully connected graph, or (ii) a sequentially connected graph, where each token is only connected to the previous tokens in the sequence. The former approach is often used for feature extraction in general (\textit{e.g., } in a Transformer-encoder), whereas the latter is employed for next-token prediction \textit{e.g.} in a Transformer-decoder). The positional information of tokens is usually encoded by positional embedding or sinusoidal positional encoding \cite{vaswani2017attention}. Transformers combine graph-like processing with the so-called attention layers. Attention layers allow Transformers to focus on ("pay attention to") the perceived relevant tokens for each prediction. Transformers have been particularly successful in sequence-to-sequence tasks, such as language translation. 
\end{itemize}

\begin{tcolorbox}[float*=th!, floatplacement=t, width=\textwidth]
\subsection*{Box 3: Structure-activity landscape modeling with geometric deep learning}

This worked example shows how geometric deep learning (GDL) can be used to interpret the structure-activity landscape learned by a trained model. Starting from a publicly available molecular dataset containing estrogen receptor binding information \cite{valsecchi2020nura}, we trained an E(3)-equivariant graph neural network (six hidden layers, 128 hidden neurons per layer) and analyzed the learned features and their relationship to ligand binding to the estrogen receptor. The figure shows an analysis of the learned molecular features (third hidden layer, analyzed \textit{via} principal component analysis; the first two principal components are shown), and how these features relate to the density of active and inactive molecules in the chemical space. The network successfully separated the molecules based on both their experimental bioactivity and their structural features (\textit{e.g.}, atom scaffolds \cite{bemis1996properties}) and might offer novel opportunities for explainable AI with GDL.\\ 

\begin{center}
\includegraphics[width=16cm]{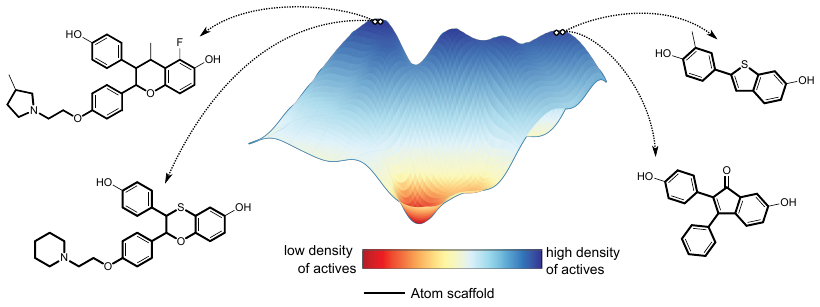}
\end{center}
\end{tcolorbox}

Extending early studies \cite{yuan2017chemspacemim,segler2018generating,bjerrum2017molecular}, RNNs for next-token prediction have been routinely applied to the \textit{de novo} generation of molecules with desired biological or physicochemical properties, in combination with transfer \cite{segler2018generating,gupta2018generative,merk2018novo,merk2018tuning} or reinforcement learning \cite{olivecrona2017molecular,popova2018deep}. In this context, RNNs have shown remarkable capability to learn the SMILES syntax \cite{segler2018generating,gupta2018generative}, and capture high-level molecular features ("semantics"), such as physicochemical \cite{segler2018generating,gupta2018generative} and biological properties \cite{merk2018novo,merk2018tuning,grisoni2021combining,yuan2017chemspacemim}. In this context, data augmentation based on SMILES randomization \cite{arus2019randomized, bjerrum2017molecular} or bidirectional learning \cite{grisoni2020bidirectional} have proven to be efficient for improving the quality of the chemical language learned by RNNs. Most published studies have used SMILES strings or derivative representations. In a few studies, one-letter amino acid sequences were employed for peptide design \cite{muller2018recurrent,nagarajan2018computational, hamid_2018_amp, grisoni2018designing,das2021accelerated}. RNNs have also been applied to predict ligand–protein interactions and the pharmacokinetic properties of drugs \cite{zheng2020predicting, wang2020optimizing}, protein secondary structure \cite{senior2019protein,zhou2020combining}, and the temporal evolution of molecular trajectories \cite{tsai2020learning}. RNNs have been applied for molecular feature extraction \cite{bombarelli_automatic_2018, lin_bigru}, showing that the learned features outperformed both traditional molecular descriptors and graph-convolution methods for virtual screening and property prediction \cite{bombarelli_automatic_2018}. The Fréchet ChemNet distance \cite{preuer2018frechet}, which is based on the physicochemical and biological features learned by an RNN model, has become the \textit{de facto} reference method to capture molecular similarity in this context.

Molecular Transformers have been applied to CASP, which can be cast as a sequence-to-sequence translation task, in which the string representations of the reactants are mapped to those of the corresponding product, or \textit{vice versa}. Since their initial applications \cite{schwaller2019molecular}, Transformers have been employed to predict multi-step syntheses \cite{schwaller2020predicting}, regio- and stereoselective reactions \cite{pesciullesi2020transfer}, enzymatic reaction outcomes \cite{kreutter2021predicting}, and reaction yields and classes \cite{schwaller2021prediction,schwaller2021mapping}. Recently, Transformers have been applied to molecular property prediction \cite{chithrananda2020chemberta, morris_2020_transformer} and optimization \cite{he2020molecular}. Transformers have also been used for \textit{de novo} molecule design by learning to translate the target protein sequence into SMILES strings of the corresponding ligands \cite{grechishnikova2021transformer}. Representations learned from SMILES strings by Transformers have shown promise for property prediction in low-data regimes \cite{honda2019smiles}. Furthermore, Transformers have recently been combined with E(3) and SE(3) equivariant layers to learn the 3D structures of proteins from their amino-acid sequence \cite{jumper2021highly,baek2021accurate}. These equivariant Transformers achieve state-of-the-art performance in protein structure prediction. 

Other deep learning approaches have relied on string-based representations for \textit{de novo} design, \textit{e.g.}, conditional generative adversarial networks \cite{mirza2014conditional, arjovsky2017wasserstein, mendez2020novo} and variational autoencoders \cite{griffiths2020constrained,alperstein2019all}. Most of these models, however, have limited or equivalent ability to automatically learn SMILES syntax, as compared to RNNs. 1D CNNs \cite{hirohara2018convolutional,kimber2018synergy} and self-attention networks \cite{zheng2019identifying,lim2020predicting,shin2019self} have been used with SMILES for property prediction. Recently, deep learning on amino acid sequences for property prediction was shown to perform \textit{on par} with approaches based on human-engineered features \cite{elabd2020amino}.

\section{Conclusions and outlook} 

Geometric deep learning in chemistry has allowed researchers to leverage the symmetries of different unstructured molecular representations, resulting in a greater flexibility and versatility of the available computational models for molecular structure generation and property prediction. Such approaches represent a valid alternative to classical chemoinformatics approaches that are based on molecular descriptors or other human-engineered features. For modeling tasks that are usually characterized by the need for highly engineered rules (\textit{e.g.}, chemical transformations for \textit{de novo} design, and reactive site specification for CASP), the benefits of GDL have been consistently shown. In published applications of GDL, each molecular representation has shown characteristic strengths and weaknesses. 

\textit{Molecular strings}, like SMILES, have proven particularly suited for generative deep learning tasks, such as \textit{de novo} design and CASP. This success may be due to the relatively easy syntax of such a chemical language, which facilitates next-token and sequence-to-sequence prediction. For molecular property prediction, SMILES strings could be limited due to their non-univocity. 

\textit{Molecular graphs} have shown particular usefulness for property prediction, partly because of their human interpretability and ease of inclusion of desired edge and node features. The incorporation of 3D information (\textit{e.g.}, with equivariant message passing) is useful for quantum chemistry related modeling, whereas in drug discovery applications, this approach has often failed to clearly outbalance the increased complexity of the model. E(3)-equivariant graph neural networks have also been applied for conformation-aware \textit{de novo} design \cite{satorras2021gen}, but prospective experimental validation studies have not yet been published. 

\textit{Molecular grids }have become the \textit{de facto} standard for 3D representations of large molecular systems, due to (i) their ability to capture information at a user-defined resolution (voxel density) and (ii) the Euclidean structure of the input grid.

Finally, \textit{molecular surfaces} are currently at the forefront of GDL. We expect many interesting applications of GDL on molecular surfaces in the near future. 

To further the application and impact of GDL in chemistry, an evaluation of the optimal trade-off between algorithmic complexity, performance, and model interpretability will be required. These aspects are crucial for reconciling the “two QSARs” \cite{fujita2016understanding} and connect computer science and chemistry communities. We encourage GDL practitioners to include aspects of interpretability in their models (\textit{e.g.}, \textit{via} XAI \cite{jimenez2020drug}) whenever possible and transparently communicate with domain experts. The feedback from domain experts will also be crucial to develop new "chemistry-aware" architectures, and further the potential of molecular GDL for concrete prospective applications.

The potential of GDL for molecular feature extraction has not yet been fully explored. Several studies have shown the benefits of learned representations compared to classical molecular descriptors, but in other cases, GDL failed to live up to its promise in terms of superior learned features. Although there are several benchmarks for evaluating machine learning models for property prediction \cite{hu2020open,wu2018moleculenet} and molecule generation \cite{polykovskiy2020molecular,brown2019guacamol}, at present, there is no such framework to enable the systematic evaluation of the usefulness of data-driven features learned by AI. Such benchmarks and systematic studies are key to obtaining an unvarnished assessment of deep representation learning. Moreover, investigating the relationships between the learned features and the physicochemical and biological properties of the input molecules will augment the interpretability and applicability of GDL, \textit{e.g.}, to modeling structure-function relationships like structure-activity landscapes (Box 3).

Compared to conventional QSAR approaches, in which the assessment of the applicability domain (\textit{i.e.}, the region of the chemical space where model predictions are considered reliable) has been routinely performed, contemporary GDL studies lack such an assessment. This systematic gap might constitute one of the limiting factors to the more widespread use of GDL approaches for prospective studies, as it could lead to unreliable predictions, \textit{e.g.}, for molecules with different mechanisms of action, functional groups, or physicochemical properties than the training data. In the future, it will be necessary to devise “geometry-aware” approaches for applicability domain assessment. 

Another opportunity will be to leverage less explored molecular representations for GDL. For instance, the electronic structure of molecules has vast potential for tasks such as CASP, molecular property prediction, and prediction of macromolecular interactions (\textit{e.g.} protein-protein interactions). Although accurate statistical and quantum mechanical simulations are computationally expensive, modern quantum machine learning models \cite{von2020exploring,unke2021machine} trained on large quantum data collections \cite{ramakrishnan2014quantum,isert2021qmugs,von2020thousands} allow quantum information to be accessed much faster with high accuracy. This aspect could enable quantum and electronic featurization of extensive molecular datasets, to be used as input molecular representations for the task of interest.

Deep learning can be applied to a multitude of biological and chemical representations. The corresponding deep neural network models have the potential to augment human creativity, paving the way for new scientific studies that were previously unfeasible. However, research has only explored the tip of the iceberg. One of the most significant catalysts for the integration of deep learning in molecular sciences may be the responsibility of academic institutions to foster interdisciplinary collaboration, communication, and education. Picking the "high hanging fruits" will only be possible with a deep understanding of both chemistry and computer science, along with out-of-the-box thinking and collaborative creativity. In such a setting, we expect molecular GDL to increase the understanding of molecular systems and biological phenomena.

\section{Acknowledgements}
This research was supported by the Swiss National Science Foundation (SNSF, grant no. 205321$\_$182176) and the ETH RETHINK initiative.

\section{Competing interest}
G.S. declares a potential financial conflict of interest as co-founder of inSili.com LLC, Zurich, and in his role as scientific consultant to the pharmaceutical industry.

\section{List of abbreviations}
\small{
\textbf{AI:} Artificial Intelligence \\
\textbf{CASP}: Computer-aided Synthesis Planning \\
\textbf{CNN}: Convolutional Neural Network \\
\textbf{DFT}: Density Functional Theory \\ 
\textbf{E(3)}: Euclidean Symmetry Group\\
\textbf{GDL}: Geometric Deep Learning\\
\textbf{GNN}: Graph Neural Network \\
\textbf{QSAR}: Quantitative Structure-Activity Relationship \\
\textbf{RNN}: Recurrent Neural Network \\
\textbf{SE(3)}: Special Euclidean Symmetry Group\\
\textbf{SMILES}: Simplified Molecular Input Line Entry Systems \\
\textbf{XAI}: Explainable Artificial Intelligence\\
\textbf{1D}: One-dimensional\\
\textbf{2D}: Two-dimensional\\
\textbf{3D}: Three-dimensional
}

\printbibliography
\end{multicols}

\end{document}